\documentclass[twocolumn,showpacs]{revtex4}
\newcommand{\gtsim}{\mbox{{\raisebox{-0.4ex}{$\stackrel{>}{{\scriptstyle\sim
}}
$}}}}
\newcommand{\ltsim}{\mbox{{\raisebox{-0.4ex}{$\stackrel{<}{{\scriptstyle\sim
}}
$}}}}

\def\cuscn{$\kappa$-(BEDT-TTF)$_2$Cu(NCS)$_2$}

\def\nh4{$\alpha$-(BEDT-TTF)$_2$NH$_4$Hg(SCN)$_4$}

\usepackage{graphicx}
\usepackage{dcolumn}
\usepackage{amsmath}
\begin{document}  
\title{Test for interlayer coherence in a
quasi-two-dimensional superconductor.}
\author{John Singleton$^{1,2}$, P.A. Goddard$^2$, A. Ardavan$^2$,
N.~Harrison$^1$, S.J. Blundell$^2$, J.A.~Schlueter$^3$ and A.M.~Kini$^3$}

\affiliation{$^1$National High Magnetic Field Laboratory, LANL, MS-E536, Los
Alamos, New Mexico 87545, USA\\
$^2$Department of Physics, University of Oxford, Clarendon
Laboratory, Parks Road, Oxford OX1 3PU, United Kingdom\\
$^3$Materials Science Division,
Argonne National Laboratory, Argonne, Illinois 60439, USA}

\begin{abstract} 
Peaks in the magnetoresistivity of
the layered superconductor \cuscn,
measured in fields $\leq 45$~T
applied within the layers,
show that the Fermi surface
is extended in the interlayer direction
and enable the interlayer transfer integral
($t_{\perp} \approx 0.04$~meV) to be deduced.
However, the quasiparticle scattering rate $\tau^{-1}$ is
such that $\hbar/\tau \sim 6t_{\perp}$,
implying that \cuscn ~meets
the criterion used to identify
interlayer incoherence.
The applicability of this criterion to anisotropic
materials is thus shown to be questionable.
\end{abstract}

\pacs{74.70.Kn, 78.20.Ls, 71.20.Rv}

\maketitle  

Many correlated-electron
systems which are of fundamental interest
have very anisotropic electronic bandstructure.
Examples include the ``high-$T_{\rm c}$'' cuprates~\cite{cuprates,strong},
layered ruthenates~\cite{ruthenate},
and crystalline organic metals~\cite{strong,review}.
Such systems may be described by a tight-binding Hamiltonian
in which the ratio of the interlayer transfer integral $t_{\perp}$
to the average intralayer transfer integral $t_{||}$ is $\ll
1$~\cite{strong,review,mck}.
The inequality $\hbar/ \tau > t_{\perp}$~\cite{mott}
where $\tau^{-1}$ is the quasiparticle
scattering rate~\cite{cuprates,strong,mck},
frequently applies to such systems, suggesting that
the quasiparticles scatter more frequently than they tunnel between layers.
The question has thus arisen as to whether the interlayer charge
transfer is coherent or incoherent,
i.e. whether or not the Fermi surface (FS) extends in the interlayer
direction~\cite{strong,review,mck}.
In this paper we have used magnetoresistance data to estimate
the interlayer transfer integral in the highly anisotropic organic
superconductor \cuscn . We find that the material obeys the
inequality $\hbar/ \tau > t_{\perp}$; moreover,
mean-free path in the interlayer direction
is $\ltsim 20$\% of the unit-cell height.
Nevertheless, our data demonstrate
a FS which is extended in the interlayer direction.

\begin{figure}[tbp]
   \centering
\includegraphics[height=8cm]{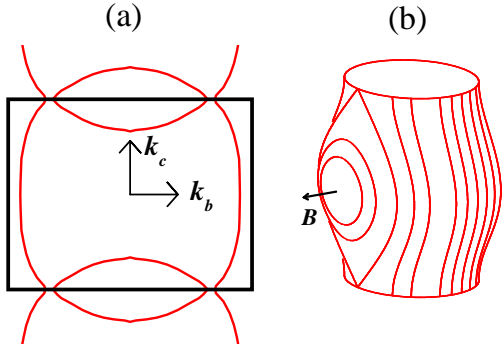}
\vspace{-2.5cm}
\caption{(a)~Cross-section of the Fermi surface (FS)
and Brillouin zone of \cuscn ~predicted by
Eqn.~\ref{phwoar}~\cite{caulfield,schmalian}.
(b)~Perspective view of the Q2D FS section described by
Eqn~\ref{phwoar}; the intralayer curvature and
interplane warping have been exaggerated for clarity.
The lines indicate
quasiparticle orbits on the FS due
to the in-plane field {\bf B}.
Note the closed orbits about the ``belly''
of the FS.}
\label{fig1}
\end{figure}

\cuscn ~was selected for our experiments because
it is perhaps the most thoroughly characterised
quasi-two-dimensional (Q2D) conductor~\cite{review}.
In contrast to the cuprates,
the FS topology is well known from Shubnikov-de Haas (SdH) and
de Haas-van Alphen (dHvA) studies~\cite{review}
and from angle-dependent magnetoresistance oscillation (AMRO)~\cite{amro}
and millimetre-wave (MMW) experiments~\cite{schrama};
it consists of a pair of quasi-one-dimensional (Q1D)
electron sheets
plus a Q2D hole pocket (see Fig.~\ref{fig1}a~\cite{caulfield,schmalian}).
The $\kappa$-phase BEDT-TTF salts are considered to be leading contenders
for interlayer
incoherence~\cite{mck}, and optical data
may be interpreted as consistent
with this suggestion~\cite{timusk}.
Moreover, models for unconventional superconductivity in $\kappa$-phase
BEDT-TTF salts invoke the nesting properties of the
FS~\cite{schmalian,charffi,aoki};
the degree of nesting might depend on whether
the FS is a 2D or 3D entity (see \cite{review},
Section 3.5).
Experimental tests for coherence in \cuscn
~are thus far deemed to be inconclusive~\cite{mck}; e.g. semiclassical
models can reproduce AMRO~\cite{amro} and MMW data~\cite{schrama}
equally well when the interlayer transport
is coherent or ``weakly coherent''~\cite{mck}.

To examine how interlayer coherence might be detected,
we use a tight-binding dispersion
relationship $-2t_{\perp}\cos(k_{\perp} a)$ to represent the interlayer
dispersion,
where $k_{\perp}$ is the interlayer component of {\bf k} and
$a$ is the interlayer spacing.
This is added to the {\it effective dimer model},
which is known to represent the
{\it intralayer} bandstructure accurately~\cite{caulfield,schmalian},
to yield
\[
E({\bf k})=
\pm 2\cos(\frac{k_{\bf b}b}{2})
\sqrt{t_{{\bf c}1}^2+t_{{\bf c}2}^2+2t_{{\bf c}1}t_{{\bf c}2}\cos(k_{\bf
c}c)}
\]
\begin{equation}
+2t_{\bf b}\cos(k_{\bf b}b)
-2t_{\perp}\cos(k_{\perp} a).
\label{phwoar}
\end{equation}
Here $k_{\bf b}$ and $k_{\bf c}$ are the intralayer components of {\bf k}
(see Fig.~\ref{fig1}a) and $t_{\bf b}$, $t_{{\bf c}1}$
and $t_{{\bf c}2}$ are interdimer transfer
integrals~\cite{caulfield,schmalian};
the $+$ and $-$ signs result in the Q1D sheets and Q2D pocket
of the FS respectively (Fig.~\ref{fig1}a).
The addition of the interlayer dispersion produces a warping of the
FS, shown schematically for the Q2D section
in Fig.~\ref{fig1}b; the FS cross-section is modulated in the interlayer
direction.
This modulation might suggest that separate
``neck and belly'' frequencies would be observed in the
dHvA effect~\cite{review}; however, only a single
frequency is seen in low-field studies~\cite{sasaki},
suggesting that the cyclotron energy
exceeds $t_{\perp}$ at the fields at which quantum oscillations are
observed.

Having failed to detect the beating between neck and bellies, and in view
of the inconclusive nature of other tests~\cite{mck}, we have chosen
instead to examine the behaviour of \cuscn ~in almost exactly in-plane
magnetic
fields~\cite{mck,hanasaki,russian}.
The motion of quasiparticles of charge $q$ in a magnetic field {\bf B} is
determined by the Lorentz force $\hbar {\bf \dot{k}}=q {\bf v} \times {\bf
B}$,
where the quasiparticle velocity is given by
$\hbar{\bf v}= \nabla_{\bf k} E$~\cite{sjb,ashcroft};
this leads to orbits on the FS in planes perpendicular to {\bf B}.
Hence, if the FS is extended in the interlayer direction, an in-plane
field will cause closed orbits on the bellies (see Fig.~\ref{fig1}b).
Such orbits are effective at averaging $v_{\perp}$,
the interlayer component of the
velocity, and their presence will lead to an increase in the
magnetoresistivity
component $\rho_{zz}$~\cite{hanasaki,russian,sjb}.
{\bf B} can then be tilted away from the in-plane direction
by an angle $\Delta$, such that
the small closed orbits about the bellies cease to be possible;
this occurs when
{\bf B} is parallel
to {\bf v} at the point at
which $v_{\perp}$ is a maximum
({\it i.e.} when the normal to the FS is
at a maximum angle to the Q2D planes).
Therefore, on tilting {\bf B} around the in-plane orientation,
we expect to see a peak in $\rho_{zz}$, of angular width $2 \Delta$,
if (and only if~\cite{mck}) the FS is extended in the interlayer direction.

A problem in using \cuscn ~is its high in-plane critical
field; $\mu_0 H_{\rm c2}(T=0) \approx 35$~T, falling to
$\mu_0H_{\rm c2}\approx 25$~T at 4.2~K~\cite{janeloff}.
Moreover, apparent peaks in the resistivity of \cuscn ~in in-plane
fields may occur due to dissipative processes associated with
vortices within the mixed state~\cite{review,chap}.
To ensure that such effects do not interfere with our data,
we choose to stay at fields well above the superconducting state;
our measurements were therefore carried out in the 45~T hybrid magnet
at NHMFL Tallahassee.
The experiments involved two
single crystals ($\sim 0.7 \times 0.5 \times 0.1$~mm$^3$; mosaic spread
$\ltsim 0.1^{\circ}$)
of $\kappa$-(BEDT-TTF)$_{2}$Cu(NCS)$_{2}$, produced using
electrocrystallization~\cite{crystal}.
In one of the crystals, the terminal hydrogens of the BEDT-TTF molecules
were isotopically substituted by deuterium;
we refer to the deuterated sample as d8, and the hydrogenated sample
as h8~\cite{footnote}.
Both crystals were mounted in a $^3$He cryostat which
allowed rotation to all possible orientations
in {\bf B}~\cite{janeloff};
sample orientation is defined by
the angle $\theta$ between
{\bf B} and the normal to the
sample's Q2D planes and the azimuthal
angle $\phi$ ($\phi=0$ is
a plane of rotation of {\bf B}
containing ${\bf k_b}$ and
the normal to the Q2D planes).
The interlayer magnetoresistance $R_{zz}$ ($\propto \rho_{zz}$)
was measured using the four-terminal ac techniques
described in \cite{janeloff}.

\begin{figure}[tbp]
   \centering
\includegraphics[height=10cm]{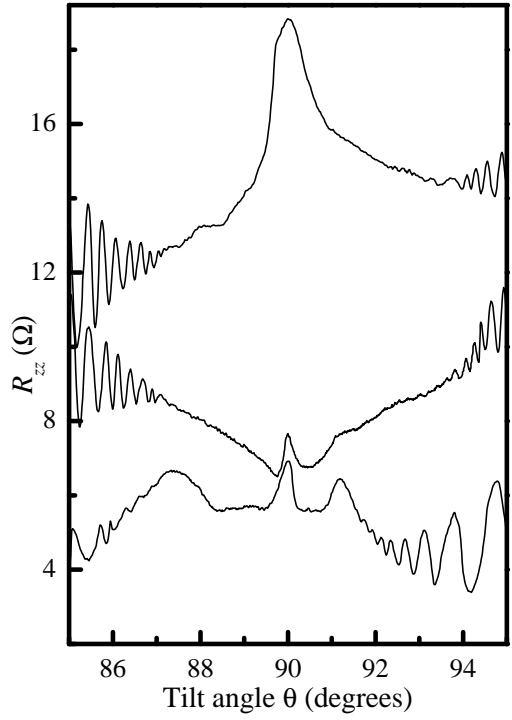}
\caption{Interlayer resistance $R_{zz}$
for the d8 \cuscn ~sample as a function
of tilt angle $\theta$.
Data for three planes of rotation of
the field are shown, $\phi=25^{\circ}$ (upper),
$\phi = 20^{\circ}$ (middle) and $\phi = 15^{\circ}$
(lower).
The static magnetic field is 42~T, and the temperature
is 520~mK.
}
\label{fig2}
\end{figure}

Fig.~\ref{fig2} shows $R_{zz}$
of the d8 sample close to the in-plane orientation $\theta =90^{\circ}$;
data for three values of $\phi$ are shown. (The data for the
h8 sample were very similar in all respects~\cite{footnote}.)
The edges of Fig.~\ref{fig2} are dominated by AMROs and related phenomena;
as these are well known~\cite{mck,sjb,goddard} in \cuscn ~\cite{amro},
we shall not describe them further in this paper.
Close to $\theta=90^{\circ}$, there is a
distinct peak in $R_{zz}$, the width and height of which vary with
$\phi$; we attribute this peak to the closed orbits described above.

\begin{figure}[tbp]
   \centering
\includegraphics[height=10cm]{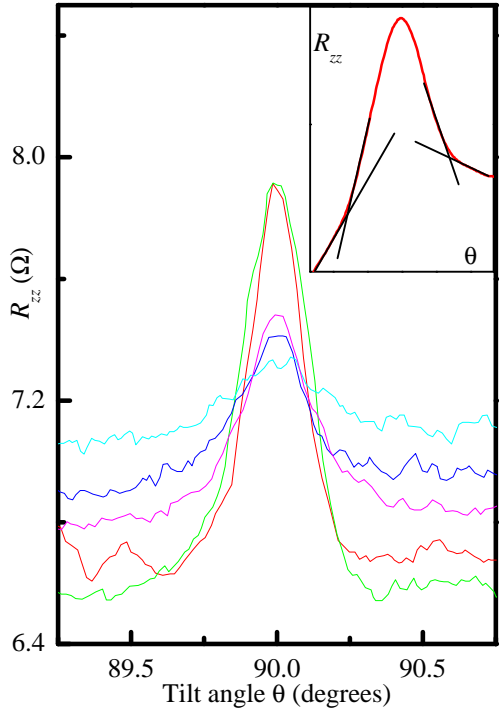}
\caption{Interlayer
resistance ($R_{zz}$) versus angle $\theta$ for temperatures
$T = 0.48$~K, 1.4~K, 3.0~K, 4.4~K and 5.1~K ($\phi=135^{\circ}$).
The background magnetoresistance increases
with increasing $T$, whereas the peak at $\theta =90^{\circ}$
becomes smaller. The data shown are for the d8 \cuscn ~sample.
The inset shows the intersections of the linear extrapolations
used to determine the peak width.
}
\label{fig4}
\end{figure}

Fig.~\ref{fig4} shows the peak at temperatures $T$ from
0.48~K to 5.1~K;
increasing $T$ by over an order of magnitude
reduces the peak height but has little effect on its width~\cite{anderson}.
Varying the field (in the range $35-45$~T
at 0.5~K, and in the range $29-45$~T at 4.2~K)
also has little effect
on the peak width; increases of field result
in increases in peak height and definition, in a manner
similar to the field dependence of AMROs~\cite{sjb}.
Both of these observations
support the idea that the peak is a consequence of the FS geometry alone.

\begin{figure}[tbp]
   \centering
\includegraphics[height=10cm]{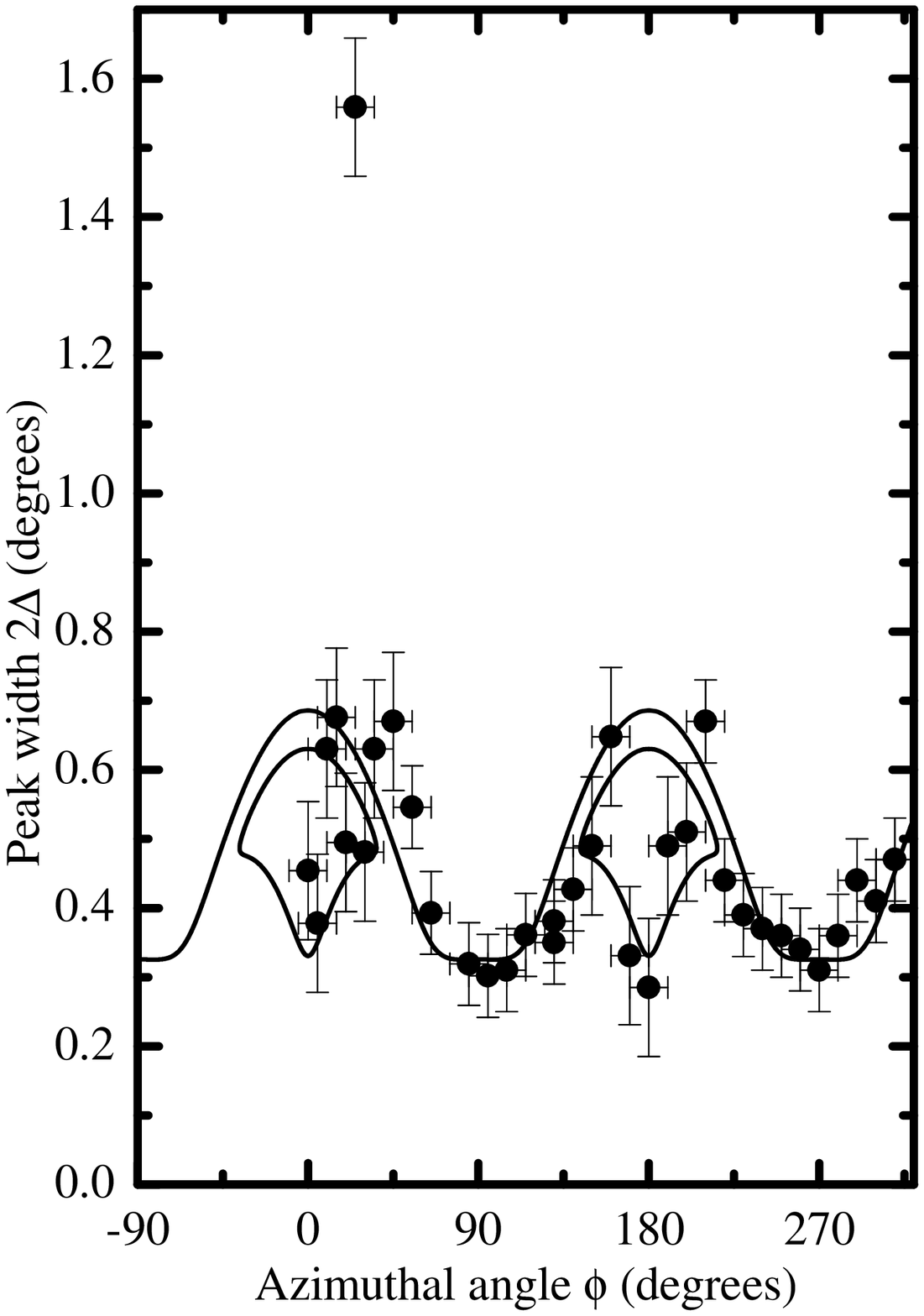}
\caption{Angular width $2\Delta$ of peak in $R_{zz}$ ($B=42$~T, $T\approx
500$~mK)
versus azimuthal angle $\phi$; data for the d8 \cuscn ~sample are shown.
Points are data; the curves represent the model prediction
with $t_{\perp}=0.04$~meV.
The continuous curve is due to the Q2D FS section; the
Q1D sheets can only support closed orbits over a restricted
range of $\phi$, leading to the top-shaped loops.
}
\label{fig3}
\end{figure}

Fig.~\ref{fig3} shows the variation of $2 \Delta$, the full width of the
peak
close to $\theta =90^{\circ}$ versus $\phi$;
the width was deduced using the extrapolations shown Fig.~\ref{fig4}
(inset).
In order to interpret this variation, we
use Eqn.~\ref{phwoar} to calculate
$v_{\perp {\rm max}}$, the maximum value of $v_{\perp}$,
and $v_{||}$, the intralayer velocity
component parallel to the plane of rotation of {\bf B}; when measured in
radians,
$\Delta \approx v_{\perp {\rm max}}/v_{||}$.
As all of the relevant quasiparticle
motion occurs close to the Fermi energy, $E \approx E_{\rm F}$, we
adjust the parameters of Eqn.~\ref{phwoar}
to reproduce the known FS parameters of \cuscn .
First, $t_{\bf c}/t_{\bf b}$ (where $t_{\bf c}$ is the mean
of $t_{{\bf c}1}$ and $t_{{\bf c}2}$) is adjusted
to obtain the dHvA frequencies of the Q2D pocket and
the magnetic breakdown orbit~\cite{caulfield,footnote,harrison}.
The absolute value of $t_{\bf c}$ is then constrained by
fitting to the effective mass of the breakdown orbit~\cite{review,footnote}.
Third, the energy gap measured in magnetic breakdown
($E_{\rm g} \approx 7.8$~meV~\cite{harrison})
gives $t_{{\bf c}1}-t_{{\bf c}2}=E_{\rm g}/2$~\cite{caulfield},
leading to $t_{\bf b}=15.6$~meV, $t_{{\bf c}1}=24.2$~meV and
$t_{{\bf c}2}=20.3$~meV.
Finally, $a=16.2$~\AA~\cite{crystal}, so that
equations for $\Delta$ contain only one adjustable parameter,
$t_{\perp}$.
The substitution of $t_{\perp}=0.04$~meV
leads to the curves shown in
Fig.~\ref{fig3}; the continuous curve is from
the Q2D FS, and the loops are caused by the Q1D sheets, which support
closed orbits only over a restricted range of $\phi$.
In the latter case, the lower branch of the loop results
from the flatter portions of the Q1D FS, roughly parallel
to ${\bf k_c}$, whereas the upper
branch is caused by the more highly curved region at the zone boundary
(Fig.~\ref{fig1}a).

For most $\phi$, there is agreement between
curve and data (Fig.~\ref{fig3}), but around $\phi=0$ and $180^{\circ}$
it seems that the observed
peak width is sometimes dominated by closed orbits on
the Q1D sheets, and sometimes by those on the Q2D FS section;
the dominant width presumably depends on which FS
section is more effective at averaging $v_{\perp}$~\cite{sjb}.
MMW studies~\cite{schrama}
suggest that the interlayer corrugations of the Q1D sheets
are more complex than those given
by Eqn.~\ref{phwoar}.
This could lead to a rapid variation with $\phi$ of the
effectiveness of the orbits in increasing $\rho_{zz}$.
For a narrow range of angles
close to $\phi=25^{\circ}$,
the peak was particularly wide and large (see top trace,
Fig.~\ref{fig2}). This is perhaps connected to
the gap between the Q1D and
Q2D FS sections.

As mentioned above,
the bandstructure of \cuscn  ~{\it within} the Q2D layers is determined
by the interdimer transfer integrals~\cite{schmalian,caulfield}.
A better guide to the {\it total} intralayer bandwidth than the
parameters used above (which are relevant for $E \approx E_{\rm F}$
and thus include renormalising interactions~\cite{review,caulfield})
is given by the fits to optical data
of Ref.~\cite{caulfield}, which suggest $t_{\bf b}\approx 60$~meV
and $t_{\bf c}\approx 120$~meV;
these are a factor $\gtsim 10^3$ larger than $t_{\perp} \approx 0.04$~meV.

The failure of the dHvA effect~\cite{sasaki}
to observe necks and bellies may now be understood; the Landau-level spacing
at
the lowest $|{\bf B}|$ used ($\sim 6$~T) is 0.2~meV, $\sim 5 t_{\perp}$.

It is instructive to compare $t_{\perp}$ with $\tau^{-1}$.
Samples d8 and h8 have been studied using SdH oscillations at
$|{\bf B}|\leq 15$~T~\cite{goddard}.
At such fields, the oscillatory
magnetoresistance is much less than
the nonoscillatory component, and magnetic breakdown
is a minor consideration~\cite{review}.
Hence, the 2D form of the Lifshitz-Kosevich
formula may be used to extract $\tau$~\cite{review},
giving $\tau = 2.9 \pm 0.5$~ps (h8) and $\tau =2.6 \pm 0.3$~ps
(d8)~\cite{goddard}.
Another estimate can be derived from MMW
studies~\cite{schrama}, which measure the
FS-traversal resonance (FTR) due to quasiparticles crossing
the Q1D FS sheets;
these experiments used samples from the same
batch as h8.
In the data of \cite{schrama},
the FTR appears
at $B \approx 10$~T, with a full-width at half-maximum of $\Delta B \approx
7$~T.
If we assume that $\omega \tau \sim B/\Delta B$~\cite{bleaney},
where $\omega = 2 \pi \times 70 \times 10^9$~rad~s$^{-1}$~\cite{schrama},
we obtain $\tau \sim 3 $~ps, close to the SdH values.
Thus, $\hbar/\tau \approx 0.24$~meV, $\sim 6 t_{\perp}$.

Finally, we estimate the intralayer mean-free path $\lambda_{||}$;
a typical value of $v_{||}$ ($\approx 7.2 \times 10^4$~ms$^{-1}$),
yields $\lambda_{||} = v_{||} \tau \approx 0.2~\mu$m
({\it i.e.} \cuscn ~is a very clean system).
However, the maximum interlayer velocity is
$v_{\perp {\rm max}} = 2t_{\perp}a/\hbar \approx 200$~ms$^{-1}$, suggesting
an
average interlayer velocity $\bar{v_{\perp}} \sim 100$~ms$^{-1}$,
and a mean-free path $\lambda_{\perp} \sim \bar{v_{\perp}} \tau \approx
3$~\AA.
The usual mean-free path formalism implies that
only $\sim \exp(-16.2/3) \approx 0.5$\% of the quasiparticles
travels between adjacent layers (separation 16.2~\AA~\cite{crystal})
without scattering.
In such circumstances, conventional criteria~\cite{mck}
lead one to expect
incoherent interlayer transport,
yet the peak in $\rho_{zz}$ described above unambiguously
demonstrates a 3D FS topology.

The criterion for incoherent transport was
developed for isotropic disordered metals~\cite{mott1};
the mean free path $\lambda$ was taken to represent
the spatial extent over which the Bloch waves are coherent and is,
by default, isotropic.
However, in Q2D
systems $\lambda_\|\gg\lambda_\bot$, so that
a purely ballistic model of transport
(of the sort used to treat disorded 3D systems~\cite{mott1})
fails to preserve information on the spatial extent of the
wavefunctions perpendicular to the layers. Our experiment therefore
demonstrates serious inadequacies in such models.
When impurities and defects are randomly dispersed throughout a material,
the inter-impurity separation should be almost isotropic.
Hence, we should also expect the spatial coherence
of the Bloch states to remain roughly isotropic,
even when the group velocity is highly anisotropic.

In summary, we observe a peak in the interlayer resistance
of the highly anisotropic superconductor \cuscn ~when
a magnetic field is applied within the layers.
This demonstrates~\cite{mck} that the
Fermi surface is extended in the interlayer direction,
and allows the interlayer transfer integral
to be estimated to be $t_{\perp} \approx 0.04$~meV ($t_{\perp}/k_{\rm B}
\approx 0.5$~K).
The small size of $t_{\perp}$ explains unsuccessful
attempts to observe neck and belly orbits in dHvA studies.
However, perhaps the most interesting point to emerge is
that \cuscn ~obeys the criteria commonly used to delineate
interlayer incoherence~- the interlayer mean free path is
smaller than the unit cell size, and $\hbar/\tau \geq t_{\perp}$~-
and yet it clearly possesses a three-dimensional Fermi surface.
It is perhaps now time to re-examine the validity of such
criteria when applied to strongly anisotropic systems.

This work is supported by EPSRC (UK).
NHMFL is supported by the
US Department of Energy (DoE), the National
Science Foundation and the State of Florida.
Work at Argonne is sponsored
by the DoE, Office of Basic Energy Sciences,
Division of Materials Science under contract number
W-31-109-ENG-38.
We thank Vic Emery
and Jane Symington
for stimulating discussions.

\begin{thebibliography}{99}
\bibitem{footnote}
AMRO and SdH data~\cite{goddard} show that
d8 and h8 have very similar FSs;
e.g. the SdH frequency
and effective mass of the Q2D pocket in
d8 (h8) are $597 \pm 1$~T ($600 \pm 1$~T)
and $3.4 \pm 0.1 m_{\rm e}$
($3.5 \pm 0.1 m_{\rm e}$)~\cite{goddard}.
The only marked difference is that the magnetic breakdown
probability is somewhat higher in d8~\cite{goddard}.
\end{thebibliography}
\end{document}